\documentclass{article}
\usepackage{spconf,amsmath,graphicx}
\usepackage{cite}
\usepackage{comment}
\usepackage{float}
\usepackage{caption, color}
\usepackage[labelformat=simple]{subcaption}

\title{Harmonicity plays a Critical role in DNN Based versus in biologically-inspired monaural speech segregation systems}
\name{Rahil Parikh$^1$, Ilya Kavalerov$^2$, Carol Espy-Wilson$^1$, Shihab Shamma$^1$ \vspace*{-3mm}}
\address{$^1$ Institute for Systems Research, University of Maryland,
	College Park, MD, USA \\
	$^2$ Google Inc, Mountain View, CA, USA
	\vspace*{-4mm}
	}
\begin{document}
%
\maketitle
\begin{abstract}
Recent advancements in deep learning have led to drastic improvements in speech segregation models. Despite their success and growing applicability, few efforts have been made to analyze the underlying principles that these networks learn to perform segregation. Here we analyze the role of harmonicity on two state-of-the-art Deep Neural Networks (DNN)-based models- Conv-TasNet and DPT-Net\cite{luo2019conv, chen2020dual}. We evaluate their performance with mixtures of natural speech versus slightly manipulated inharmonic speech, where harmonics are slightly frequency jittered. We find that performance deteriorates significantly if one source is even slightly harmonically jittered, e.g., an imperceptible $3\%$ harmonic jitter degrades performance of Conv-TasNet from 15.4 dB to 0.70 dB.
Training the model on inharmonic speech does not remedy this sensitivity, instead resulting in worse performance on natural speech mixtures, making inharmonicity a powerful adversarial factor in DNN models. Furthermore, additional analyses reveal that DNN algorithms deviate markedly from biologically inspired algorithms\cite{krishnan2014segregating} that rely primarily on timing cues and not harmonicity to segregate speech.

\end{abstract}
\begin{keywords}
speech segregation, Conv-TasNet, harmonic structure, temporal coherence, adversarial inputs
\end{keywords}
\section{Introduction}
\label{sec:intro}
The `cocktail party problem' is the task of attending to a source of interest, usually speech, in a complex acoustic environment with concurrent sounds. Despite the apparent ease with which humans can group acoustic cues from such an environment and organize them to meaningfully perceive them \cite{bregman1994auditory}, the complexity of this problem has inspired generations of neuroscientists, psychologists and engineers. The multi-disciplinary nature of this problem has motivated Computational Auditory Scene Analysis (CASA) algorithms using traditional pitch modulations extractions \cite{vishnubhotla2009algorithm,stark2010source, wang1999separation} and biologically-inspired systems which are based on frameworks such as `temporal coherence' (TC) \cite{elhilali2008cocktail, krishnan2014segregating}. In contrast to these approaches, there are strictly engineering frameworks such as signal factorization, statistical-speech modelling \cite{ weiss2010speech, cooke2010monaural} and recently deep learning \cite{huang2015joint, hershey2016deep, zhang2016deep, luo2018tasnet, luo2019conv, luo2020dual, chen2020dual, kolbaek2017multitalker} which ignore biological plausibility. 
Advances in deep learning have facilitated drastic improvements in speech segregation \cite{ hershey2016deep, zhang2016deep, luo2018tasnet, luo2019conv, luo2020dual, chen2020dual, kolbaek2017multitalker}. These models can now be used at scale, to separate individual speech from mixtures, enabling downstream audio processing tasks such as Automatic Speech Recognition (ASR), allowing us to use speech technologies in the \textit{real-world}. These model architectures have also shown promising results in music segregation \cite{defossez2019music} and sound segregation \cite{kavalerov2019universal}. Despite their unquestionable success, no work has been done to investigate the underlying principles they learn to perform grouping and consequently segregation. 

We attempt to bridge the gap between our understanding of traditional CASA systems, biologically inspired algorithms and deep-learning systems by investigating the role of harmonicity on the latter. Since harmonicity of natural speech is linked with pitch extraction, it is a crucial cue for grouping in certain CASA systems \cite{vishnubhotla2009algorithm, wang1999separation}. However, biologically plausible algorithms based on TC depend solely on the temporal correlation of features from the same source and are capable of performing segregation regardless of harmonicity \cite{krishnan2014segregating}. While recent findings in \cite{popham2018inharmonic} found that human subjects perform worse in segregating inharmonic acoustic mixtures, performance deficits were relatively modest and often attributable to source assignment confusion rather than true segregation difficulty of the overlapping words in the mixtures. 
In our work, we begin by exploring the importance of harmonicity on two prominent speech segregation models Conv-TasNet \cite{luo2019conv} and Dual-Path Transformers (DTP) Net \cite{chen2020dual}. We investigate how these models perform on mixtures of harmonic speech, inharmonic speech, and a variety of synthetic simple stimuli that exemplify these properties more clearly. We also extend this analysis to the same models when trained on inharmonic speech. We find that- 
\begin{itemize}
    \item DNN based segregation systems trained on natural speech  fail to segregate inharmonic speech. The performance of DPT-Net degrades from 20.2 dB to 0.7 dB for imperceptible amounts of harmonic jitter.
    \item DNNs find it hard to learn to segregate sources when trained on inharmonic speech, instead becoming  worse at segregating natural speech.
    \item They differ from biologically inspired systems in heavily relying on harmonic patterns to segregate speech.
\end{itemize}


\section{Background}
\subsection{Monaural Speech Segregation Models}
\label{sec:Monaural Speech Separation Models}
Monaural speech segregation involves estimating $C$ sources $s_{1}(t), \cdots, s_{C}(t)$ for an audio mixture $x_{mix}(t) = \sum_{i=1}^{C}{s_{i}}$. In this work we focus on end-to-end speech segregation models \cite{luo2018tasnet} which share a similar pipeline-- the input mixture is encoded using a learnt analysis transform. The segregation network generates $C$ masks for this encoded input. The masked input is then decoded using a learnt synthesis transform to generate a time-domain estimate for each source. These networks typically use Permutation Invariant Training (PIT) \cite{kolbaek2017multitalker} to maximize the Scale Invariant Source-to-Noise Ratio (SI-SNR) \cite{luo2018tasnet}.
In Conv-TasNet the segregation network consists of dilated 1-D convolutional networks, whereas  DPTNet consists of a cascade of transformers configured to maximize the temporal receptive field over the encoded mixture. In this work we set $C=2$ for simplicity and illustrative purposes.

\subsection{Inharmonic Sound Sources}
\label{Inharmonic Sound Source}
Sounds that originate from a periodic process have a harmonic structure in their Fourier representation, consisting usually of a component at the fundamental frequency (F0) and its integer multiples.
Inharmonic sounds instead do not have frequency components at exact integer multiples of F0. We use the modified STRAIGHT algorithm \cite{ellis2012inharmonic} to generate inharmonic speech. This algorithm models voiced excitations as a sum of multiple time varying sinusoids, the frequencies of which are manipulated using `jittering', wherein a random offset is added to each harmonic. This offset is bounded by some fraction- $J$ of F0. The $n^{th}$ harmonic is thus modified to:
\vspace*{-2mm}
\begin{equation}
    f_{n}(t) = n f_{0}(t) + J_{n} f_{0}(t) ;~ -J \le J_{n} \le J
\label{eqn: jittered_freq}
\vspace*{-2mm}
\end{equation}
Since other characteristics of the speech remain unchanged, synthetically generated inharmonic speech is very intelligible\footnote{http://labrosa.ee.columbia.edu/projects/inharmonic/}. 
To generate inharmonic tone complexes, we sample from $x_{tone}(t) = \sum_{k=1}^{N} A_{k} sin(2\pi f_{n}(t) t)$
where $f_{n}(t)$ is given by (\ref{eqn: jittered_freq}). Sources are harmonic at $J_{n}=0$.

\section{Experiments}
\label{sec:Experiments}
To demonstrate the importance of harmonicity as a cue for grouping, we first evaluate performance of Conv-TasNet (trained on native (harmonic) WSJ-2-mix) with mixtures of harmonic or inharmonic complex tones. To extend these results to speech, we evaluate Conv-TasNet and DPT-Net with inharmonic versions of WSJ-2-mix, with varying levels of inharmonicity. We also evaluate these models with mixtures where one speaker is natural and the other is inharmonic. Finally, we train a Conv-TasNet on only inharmonic mixtures.

\subsection{Dataset Construction and Evaluation} \label{sec:Datasets}
We use the native WSJ-2-mix dataset \cite{hershey2016deep} created using the WSJ0 corpus to train our models on harmonic speech. We create multiple inharmonic versions of the WSJ0-corpus, by varying the harmonic jitter $J$ in (\ref{eqn: jittered_freq}) and create their corresponding inharmonic WSJ-2-mix datasets. We change $J$ from $0.01-0.30$, corresponding to an average offset in the range of $\pm1.2-\pm40$ Hz for male speakers and $\pm2.1-\pm65$ Hz for female speakers. A higher value of $J$ corresponds to a higher deviation from the native source. We also generate versions of the WSJ-2-mix datasets wherein one source is harmonic and the other is inharmonic. 

We evaluate source-segregation performance using the metric signal-to-distortion-ratio improvement (SDRi) \cite{vincent2006performance}. The results are baselined against the model's performance on the harmonic WSJ-2-mix dataset with $J=0$. Since each of the speech datasets contain differently jittered versions of the same samples from WSJ-2-mix, we can fairly compare the evaluation performance across these datasets.



\subsection{Results} \label{sec:Results}
\subsubsection{Speech Segregation Models Trained on Natural Speech}
\begin{figure}[!htbp]
  \centering
  \includegraphics[width=0.9\columnwidth]{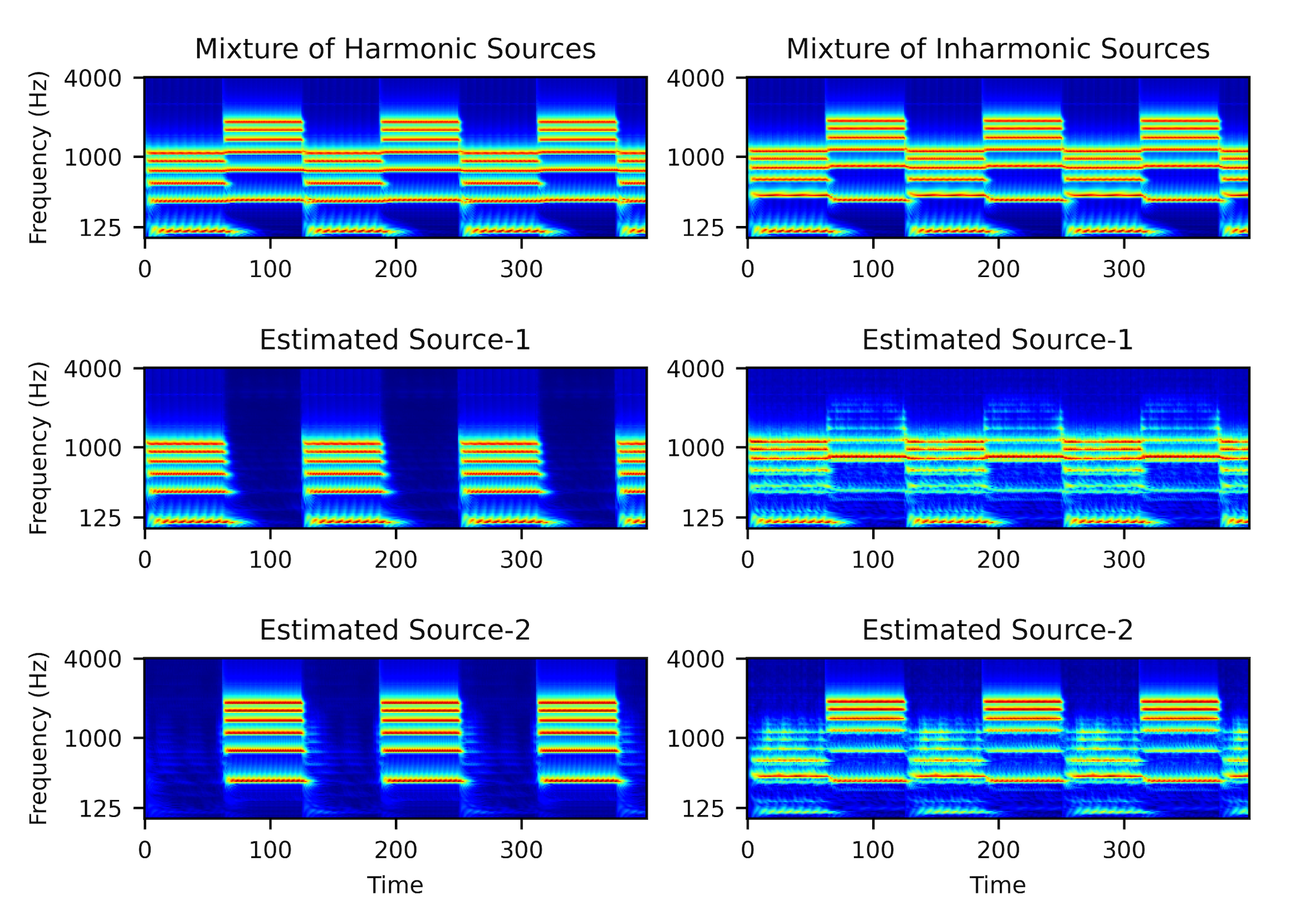}
  \caption{Conv-Tasnet Segregates mixtures of harmonic tone complexes (left panels), but  fails with inharmonic complexes (right panels).}
  \label{fig:plot_tones_harmonic_inharmonic}
\end{figure}

We begin by demonstrating that Conv-TasNet trained on the native WSJ-2-mix relies primarily on harmonicty to perform separation. The test stimuli are mixtures of alternating tone complexes (with no overlap) as illustrated in Fig \ref{fig:plot_tones_harmonic_inharmonic}.
The model performs well when both complexes are harmonic ($J =0$ and $F0=110, 210$ Hz) as illustrated in the left panels of Fig. \ref{fig:plot_tones_harmonic_inharmonic}. It, however, fails completely  when the complexes are harmonically-jittered (right panels) with $J = 0.20$. 

\begin{figure}[!htbp]
  \centering
  \includegraphics[width=0.9\columnwidth]{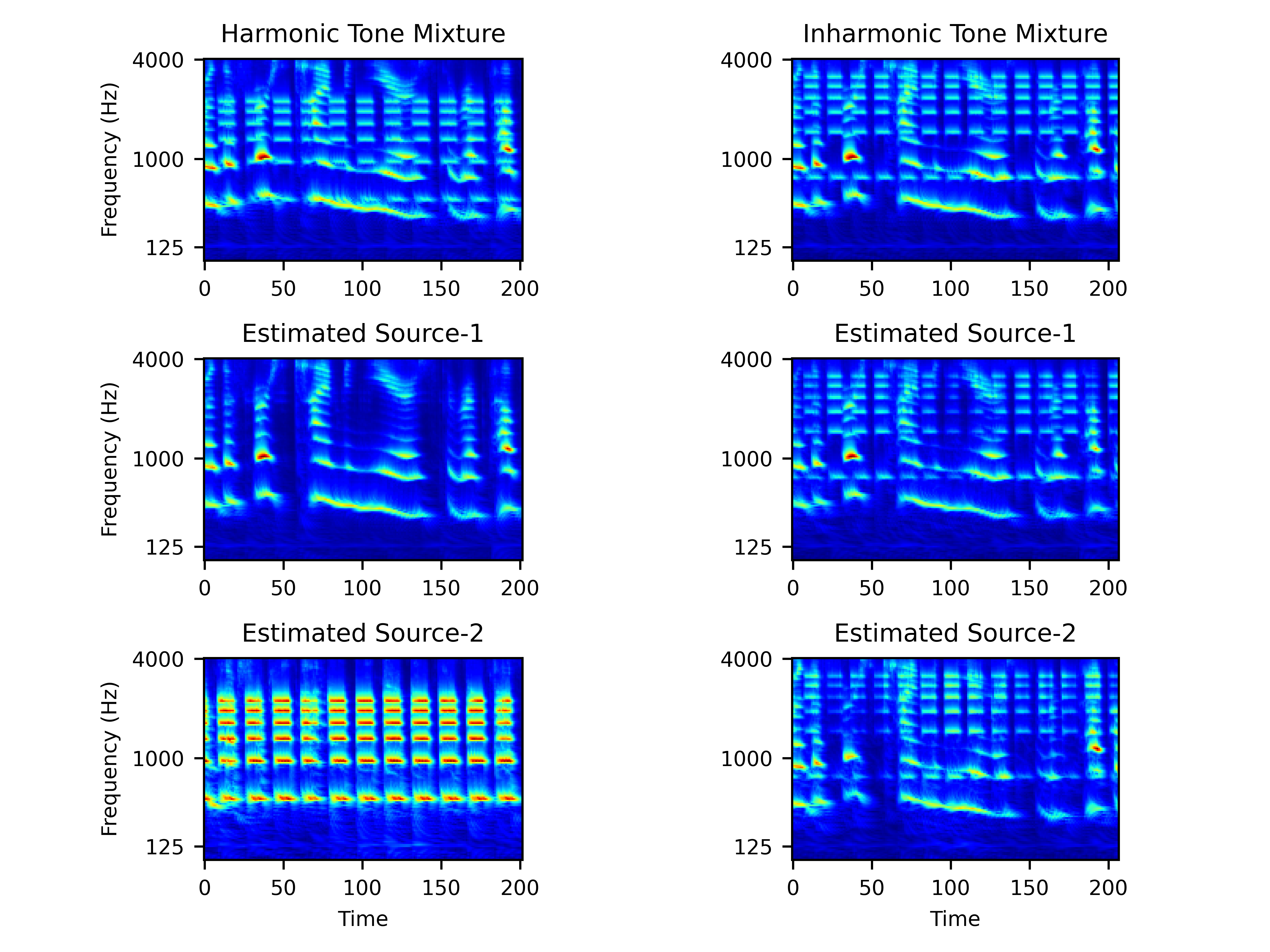}
  \caption{Mixtures of natural speech with harmonic (left panels) and inharmonic (right panels) tone complexes are segregated only when both are harmonic.}
  \label{fig:plot_speech_and_tones_mix}
\end{figure}

This phenomenon applies equally when the pair of stimuli are very different as in Fig.\ref{fig:plot_speech_and_tones_mix}, where one is a tone complex and the other is speech. The segregation is perfectly effective when both are harmonic (left panels) while it fails when one of the sources (the tone complex) is inharmonic (right panels) in spite of the speech being harmonic.

\begin{figure}[htp]
  \centering
  \includegraphics[width=0.85\columnwidth]{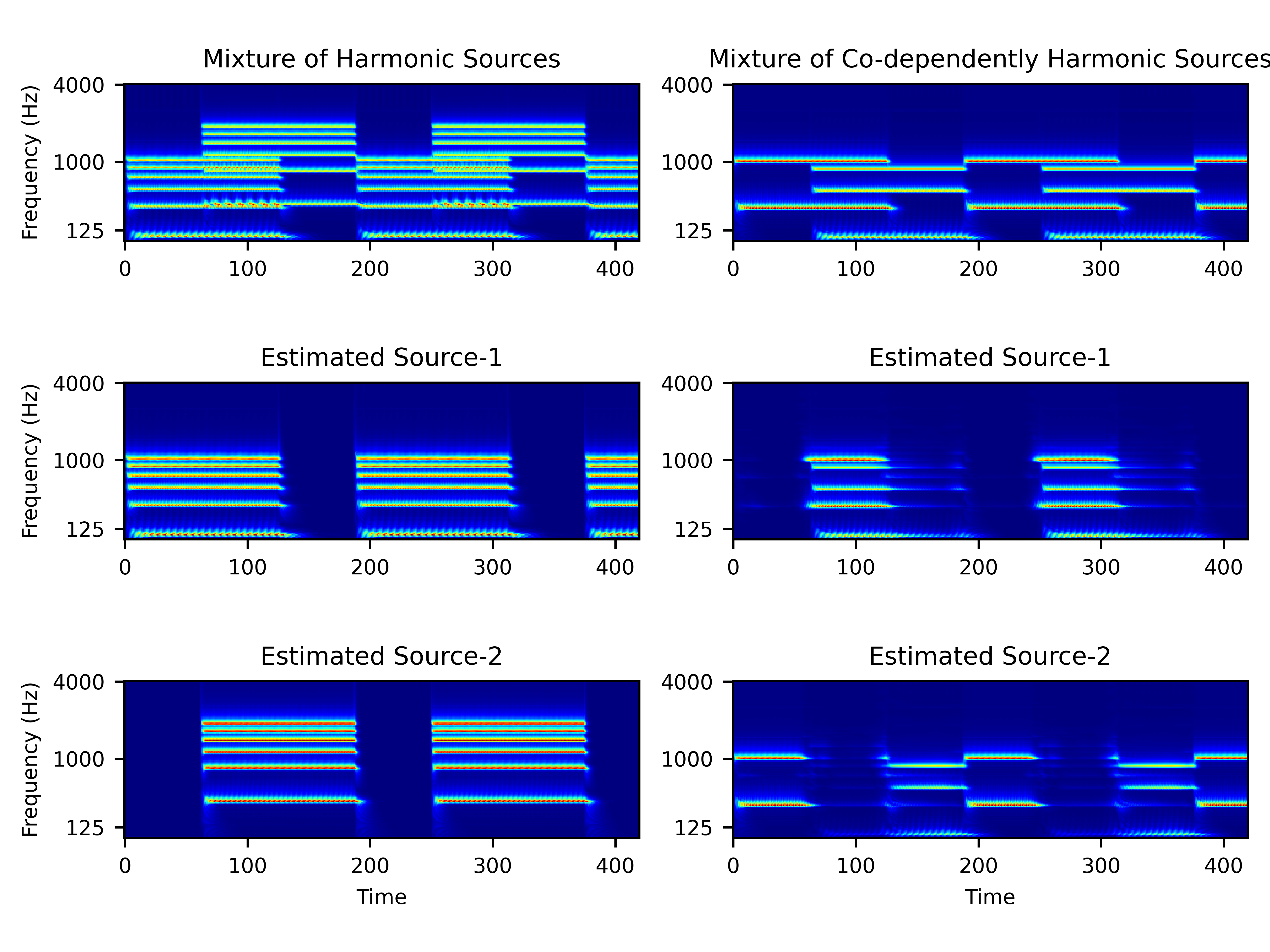}
  \caption{Segregation of overlapped spectra. (Left panels) A mixture of 2 harmonic sources segregate well. (Right panels) Alternating inharmonic spectra  become harmonic and incorrectly segregate during overlap.}
  \vspace*{-3mm}
  \label{fig:overlap_harmonics}
\end{figure}

\begin{figure}[t]
  \centering
  \begin{subfigure}[l]{0.23\textwidth}
  \includegraphics[width=1.7in]{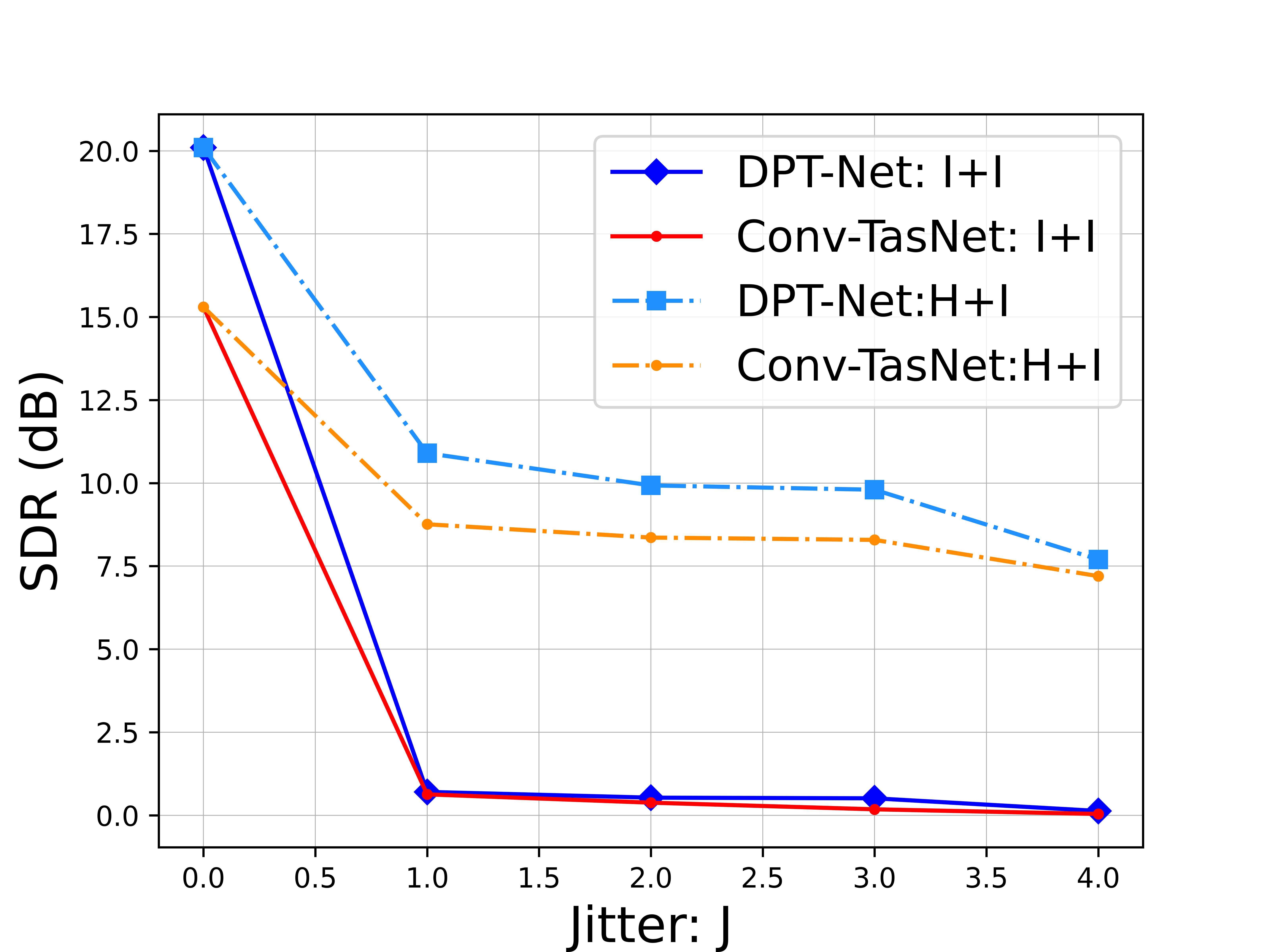}
  \caption{}
  \label{fig:plot_trained_net}
  \end{subfigure}
\begin{subfigure}[l]{0.23\textwidth}
  \includegraphics[width=1.7in]{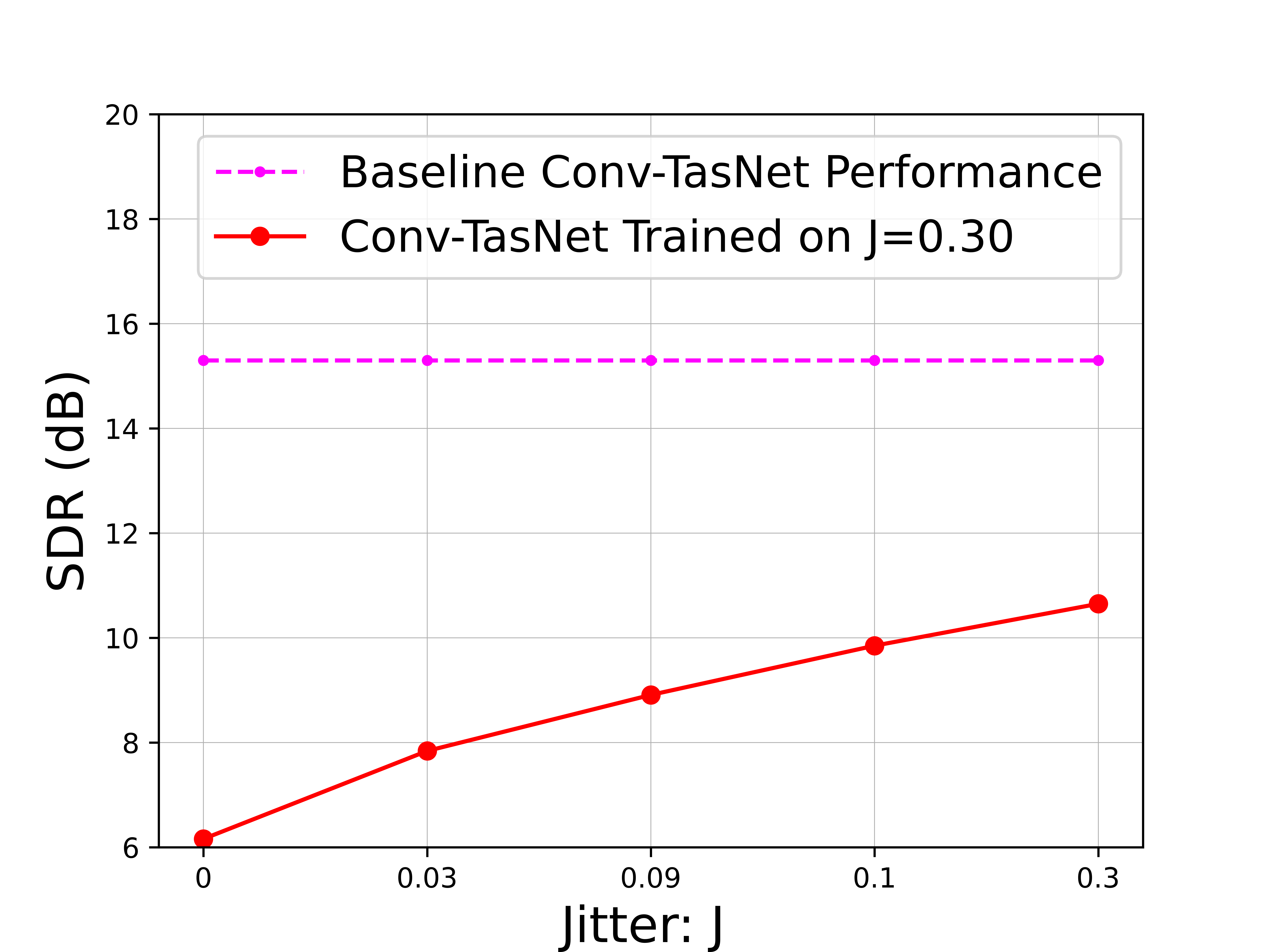}
  \caption{}
  \label{fig:plot_train_on_inharmonic_only}
  \end{subfigure}
  \caption{Performance on mixtures of inharmonic speakers with increasing jitter $J$ for models trained on (a) native WSJ-2-mix and (b) inharmonic WSJ-2-mix with $J=0.3$}
 \vspace*{-6mm}
\end{figure}

\vspace*{-2mm}
Harmonicity may also be a transient phenomenon during speech mixtures due to the overlap of harmonic components. To illustrate the immediate sensitivity of the DNN models to the emergence (or absence) of harmonicity, we illustrate in Fig.\ref{fig:overlap_harmonics} two alternating mixtures with  50\% overlap. In the left panels, both tone complexes are completely harmonic (F0=$100$ and $190$). On the right, the  complexes are inharmonic with frequencies  $200, 600$ Hz and $100, 300, 500$ Hz. During the intervals of overlap, the effective instantaneous spectrum becomes the sum of the two complexes. On the left, the harmonic patterns still exist in the mixture, and hence are extracted out and segregated perfectly. But in the right panels, the tone mixture strongly resemble the harmonics of F0=$100$ Hz. The DNN model incorrectly groups this harmonic series of tones as one source, while the two individually inharmonic series become the other source. This demonstrates the strong dependence of the model on the harmonicity of the spectrum.\\
These tonal stimuli illustrate the vulnerabilities that affect  models when harmonicity is disturbed. To extend our analysis to speech mixtures we evaluate the models using mixtures of: two inharmonic speakers (I+I), and one harmonic and one inharmonic speaker (H+I), using datasets discussed in Sec. \ref{sec:Datasets}. The results in Fig.\ref{fig:plot_trained_net} illustrate that the models' performance degrades severely, especially on mixtures of two inharmonic speakers-- 15 dB on natural speech to 0.7 dB for ConvTasnet and 20 dB to 0.7 dB for DPT-Net, for $J=0.03$, which reveals a remarkable vulnerability.
\vspace*{-3mm}

\subsubsection{Training DNN-Models with Inharmonic Speech}
\vspace*{-1mm}
An important question is whether these DNN models can be made more robust to inharmonicity by training them on inharmonic speech. Here, we trained the Conv-TasNet on the WSJ-2-mix datasets with both inharmonic speakers. We note that the model finds it more challenging to learn to segregate with increasing inharmonicty. Our results from a model trained only on inharmonic data with $J=30\%$ are illustrated in Fig. \ref{fig:plot_train_on_inharmonic_only}. In spite of being trained on data from the same distribution, the performance of the network drops from 15.3 dB to 10.6 dB. Interestingly, this network now performs poorly on harmonic and mildly inharmonic speech. Furthermore, when the model is trained on a combination of natural and inharmonic mixtures, it still finds it hard to segregate inharmonic mixtures but learns to segregate natural speech. We believe that learning to segregate inharmonic mixtures is a harder problem since the space of inharmonic patterns is very large, while that of harmonic patterns is sparse. This makes it difficult to cover all possibilities during training. These results underscore that the networks rely strongly on the constrained harmonic structure for segregation.

\vspace*{-1mm}

\subsection{Discussion} \label{Discussion}
\begin{figure}[!htp]
  \vspace*{-2mm}
  \centering
  \includegraphics[width=0.90\columnwidth]{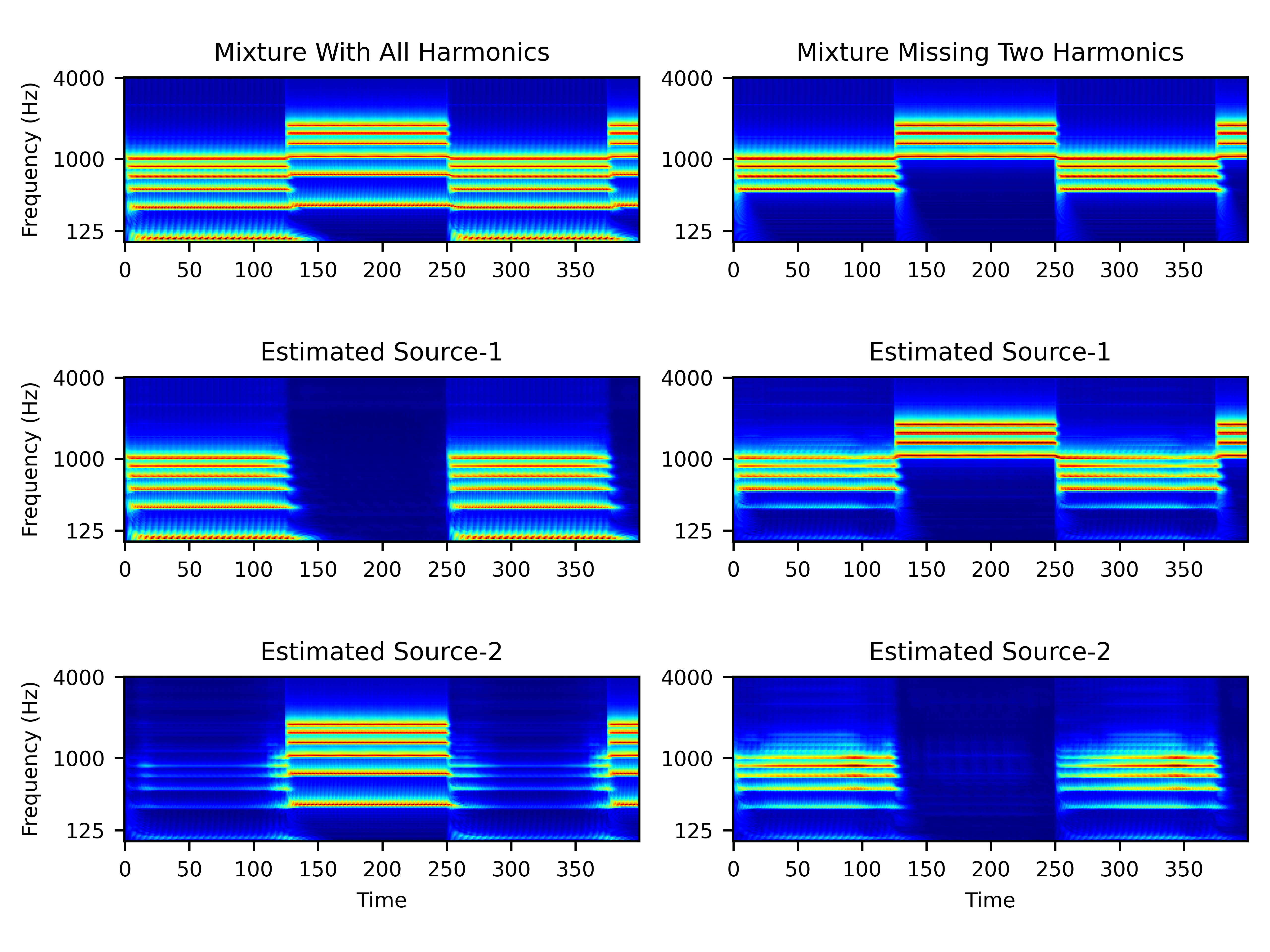}
  \caption{Conv-Tasnet successfully segregates mixtures of harmonic sources (left panels). However, it fails when the fundamentals and some lower harmonics are absent (right panels).}
  \label{fig:missing_harmonics}
  \vspace*{-7mm}
\end{figure}

Our experiments indicate that end-to-end models trained on natural speech are incapable of segregating an inharmonic source. Furthermore, we have shown that the models cannot even be trained to segregate inharmonic mixtures.\\ These results provide clear evidence that DNN-based speech segregation is heavily reliant on learning the harmonic patterns that are available in natural speech, but that are absent or rarely available in inharmonic training. Their sensitivity to harmonicity implies the ability to track the F0, or pitch. But they seem to do so from the full spectral set of harmonic patterns and not just from a few harmonics as humans are readily able to do. For instance, in Fig.\ref{fig:overlap_harmonics}, the network could not treat the sparser pattern of tones as part of a larger harmonic pattern, expecting instead to see the full patterns. 
To further test this, we generate a mixture of alternating harmonic tone complexes with and without the fundamental component and the second harmonic, as illustrated in Fig. \ref{fig:missing_harmonics}. Since the model fails to segregate the tones, it is likely it has learned to recognize only the full harmonic patterns. They then proceed to \textit{track} the trajectory of the pitches implied by these harmonics the varied mixtures of speakers and tones, and despite their often proximate values, patterns, and even inter-crossings and gaps and other well-known difficulties. These experiments provide evidence of the important cues involved in the remarkable performance of the DNNs, but they also demonstrate their vulnerabilities. \\

\vspace*{-5mm}

In additional experiments not described here for the sake of brevity, we also observed that the models have difficulties in segregating fricatives and unvoiced segments of speech. Fricatives are plentiful in the training set but are not harmonic in structure, indicating that these vulnerabilities are not a result of testing on `out-of-distribution' data, but instead  stem from the underlying mechanisms used by the DNNs to perform the segregation. Finally, we have also confirmed that these same vulnerabilities are not limited to time domain based models trained in an end-to-end fashion but also extend to spectrogram based segregation models.

\subsubsection{Divergence of DNN Models from Temporal Coherence} 
\begin{figure}[]
  \centering
  \includegraphics[width=\columnwidth]{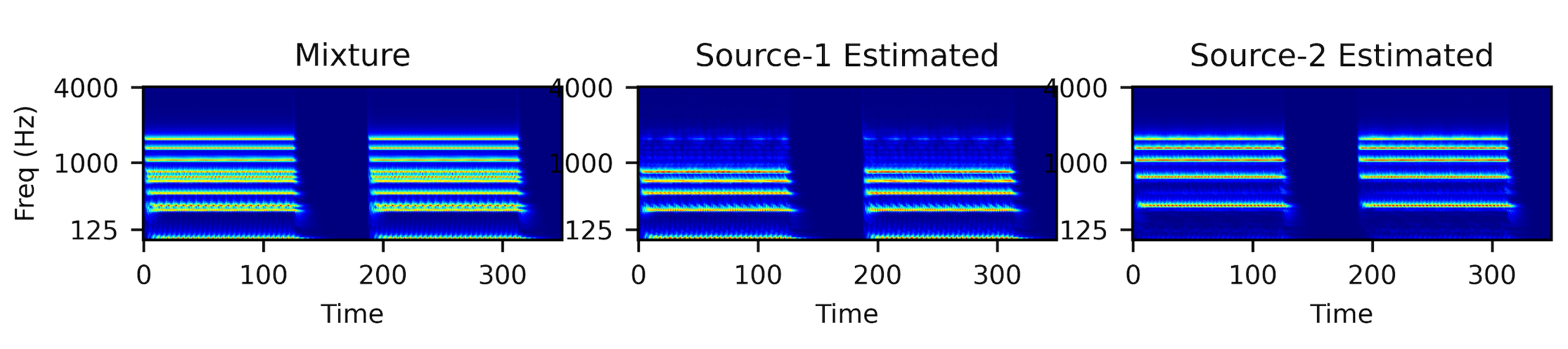}
  \caption{Conv-TasNet segregates two harmonic sources even when synchronous.}
  \vspace*{-4mm}
  \label{fig:asynchronous sources}
\end{figure}

\vspace*{-1mm}
Our results also indicate another significant  divergence of the DNN-based segregation models from biologically plausible and perceptual models that perform speech segregation using TC. While TC partially leverages pitch information and consequently harmonicity, they primarily rely on the coincidence of the onsets and timing information to group the features emanating from a single source. This divergence between DNNs and such algorithms is evident in Fig. \ref{fig:asynchronous sources}, which illustrates that Conv-TasNet can generally segregate two harmonic complexes that have the same exact onset and offset time. Humans and TC algorithms fail to do so because they would group all these components as coming form a single source. Conversely, unlike Conv-TasNet, TC algorithms would have been able to separate the tones in Fig. \ref{fig:overlap_harmonics}.
\vspace*{-1mm}
\vspace*{-2mm}

\subsubsection{Inharmonic Sources as Adversarial Inputs}
\vspace*{-1mm}
We have demonstrated that speech segregation networks perform poorly if an inharmonic speech or tone source is present in the mixture. An adversary can thus generate inharmonic speech which sounds similar to natural speech as an adversarial input to hinder source-segregation and its downstream-tasks. Since the presence of an inharmonic source is independent of the model architecture, the adversarial nature of such sources is not limited to Conv-TasNet and DPT-Net. 

\vspace*{-2mm}
\section{Conclusion}
\vspace*{-1mm}
We have shown that DNN-based speech segregation models such as Conv-TasNet and DPT-Net rely on the harmonicity of the sources in the mixture to separate them. They, therefore,  fail to segregate sources when their harmonic structure is disturbed, and also  find it harder to learn to segregate mixtures with such inharmonic sources. This study  demonstrates that these networks implicitly learn to \textit{estimate} pitch by learning the harmonic patterns abundant in the training corpus. They then proceed to track these harmonic complexes to segregate speech. This effort to understand the underlying principles that these networks learn helps us narrow the gap between our understanding of auditory perception and end-to-end learning-based segregation models. 

\vspace*{-1mm}
\section{ACKNOWLEDGEMENTS}
\label{sec:typestyle}
\vspace*{-2mm}
This work was supported by NSF grant \#1764010 and an AFOSR grant. The authors declare no conflict of interests.
%


\vfill\pagebreak



\bibliographystyle{IEEEbib}
\bibliography{strings,refs}

\end{document}